\documentclass[hyper]{JHEP} 

\usepackage{epsfig}

\newcommand\fverb{\setbox\pippobox=\hbox\bgroup\verb}
\newcommand\fverbdo{\egroup\medskip\noindent%
			\fbox{\unhbox\pippobox}\ }
\newcommand\fverbit{\egroup\item[\fbox{\unhbox\pippobox}]}
\newbox\pippobox
\title{Note About   Non-BPS and BPS Dp-branes
in Near Horizon Region of $N$  Dk-Branes}

\author{by J. Kluso\v{n}\\
	 Department of Theoretical Physics and Astrophysics\\
                   Faculty of Science, Masaryk University\\
Kotl\'{a}\v{r}sk\'{a} 2, 611 37, Brno\\
Czech Republic\\
	E-mail: \email{klu@physics.muni.cz}}

\preprint{\hepth{0502079}}

\abstract{In this paper we will consider  the dynamics
of BPS and non-BPS Dp-branes in the background
of $N$ Dk-branes. Our approach 
is based on an existence of
 the new symmetry
of  D-brane effective actions  
that naturally
emerges in the near horizon region of
the stack of  $N$ Dk-branes. 
 Since generally this
scaling symmetry is explicitly broken
in the Lagrangian we will find
 the equation that determines the
time evolution of the generator of this
transformations.  
Then we will argue that
in case when the tachyon  living on
the worldvolume of unstable D-brane
reaches the  stable minimum 
the time evolution of this generator can be  easily
determined.  With the help of 
 the knowledge of the   time dependence of this charge
we will 
determine the trajectory of the non-BPS D-brane in the
near horizon region of $N$ Dk-branes.
In case of BPS Dp-brane probe we will aruge
that such a broken scaling symmetry exists as
well and the existence of the explicit
time dependence of the generator of this symmetry
can be used in the solving the equation of motion
of the probe Dp-brane in the near horizon  
region of $N$ Dk-branes.}

\keywords{D-branes}

\begin{document}
\section{Introduction}\label{first}
The study of unstable objects
in string theory 
\footnote{For recent review that
also contains extensive list of
references, see \cite{Sen:2004nf}.}
led to the deep insight into
the string theory. As is well know
the decay of unstable D-brane is
triggered by  rolling the open
string tachyon 
\cite{Sen:2002qa,Sen:2002in,Sen:2002nu}.
An interesting consequence of
this research is the fact that
the effective field theory description
of the tachyon condensation
\cite{Sen:1999md,Kluson:2000iy,Bergshoeff:2000dq,
Garousi:2000tr}
 is very powerful even if the tachyon
is of order of the string scale.
This issue was extensively discussed
in \cite{Kutasov:2003er,Niarchos:2004rw,Sen:2003zf,
Fotopoulos:2003yt}.

A situation in which dynamics
involved is quite analogous to the
tachyon rolling is the formation
of non-threshold bound states. 
A particular example of such a 
process is falling probe BPS D-brane
into a target stack of $N$ NS5-branes
\cite{Kutasov:2004dj}
\footnote{Related problems were also
studied in 
\cite{Nakayama:2004ge,Chen:2004vw,Thomas:2004cd,
Bak:2004tp,
Saremi:2004yd,Kutasov:2004ct,Sahakyan:2004cq,
Ghodsi:2004wn,Panigrahi:2004qr,Yavartanoo:2004wb,
Nakayama:2004yx,Kluson:2005qx}.}.
Interestingly, the dynamics
of the rolling of the radial 
mode(``radion'') resembles tachyon
rolling dynamics of unstable D-brane. 
In particular, for appropriate background
and regime of target NS5-branes
or Dk-branes the works cited above
noted that the radion effective action
takes exactly the same functional
form as the tachyon effective action
for unstable D-brane. The natural 
proposal is then to view the radion rolling
dynamics as a sort of ``geometrical
realisation'' of the tachyon
rolling dynamics for unstable
D-brane. 

The extension of these works to the
case of a 
non-BPS Dp-brane probe has been
done in  
\cite{Kluson:2004yk,Kluson:2004xc,Kluson:2005qx}.
Generally the study of the dynamics
of non-BPS Dp-brane is much more involved
than in case of BPS Dp-brane thanks to
the presence of the tachyon on the
worldvolume of non-BPS Dp-brane. However
we have argued in
\cite{Kluson:2004yk,Kluson:2005qx}
that in region of the
tachyon and radion field theory
space where tachyon is large and 
Dp-brane is in the near horizon region
of $N$ Dk-branes of NS5-branes
there exists   additional
worldvolume 
symmetry 
\footnote{The form of this symmetry 
is similar to the symmetry discussed in 
\cite{Jevicki:1998ub,Jevicki:1998yr}.}
of the action that simplifies the
analysis of the resulting
equations of motion and
we were able to study 
the time evolution of the radion
and tachyon on the worldvolume 
of a non-BPS Dp-brane.
It is however important to stress
that the existence of this symmetry
was restricted to the case of
the motion of Dp-brane in the
background of $N$ NS5-branes or
in the background of $N$ D3-branes. 

In this   paper we will 
continue further the study of the
properties of this new
symmetry (possible explicitly broken)
of the tachyon effective action
in the Dk-brane background. We will find
that the tachyon effective
action in the near horizon region
and for large tachyon is
invariant under the form of
scaling transformation introduced
bellow on condition that the
parameter $\lambda\equiv Ng_sl_s^{7-k}$,
where $N$ is a number of background
Dk-branes, $l_s$ is the string
length and $g_s$ is string coupling,
 transforms 
under these transformations as well.
We will 
 determine the generator
of the scaling symmetry from
the tachyon effective action and
then  we  find the equation
that governs the time evolution
of the generator $D$ 
\footnote{We label by symbol $D$ the generator
of the scaling symmetry.}.   
It turns out however that
for the tachyon that reaches
its stable minimum 
 this equation 
 takes particular
simple form  that allows us to 
find the explicit time dependence of $D$. 
Then using this fact
 we will solve 
the equations of motion
for $R$ and hence we will able to
 determine 
the time evolution of the non-BPS Dp-brane
in the near horizon region of
$N$ Dk-branes in situation 
when 
$T$ reaches its stable minimum.

The paper is organised as follows.
In the next section (\ref{second})
we introduce the  form
of the tachyon effective action 
in the near horizon region of $N$
Dk-branes.
Then we will study the symmetries
of this action  using careful
application  of the Noether theorem.
 In section
(\ref{third}) we will study the
time evolution of the non-BPS Dp-brane
where the worldvolume tachyon is
in its stable minimum. 
In section (\ref{fourth})
we will study 
 the dynamics of a BPS Dp-brane in the
background of $N$ Dk-branes 
that is based on an existence
of the broken scaling symmetry. 
Finally, in conclusion (\ref{fifth})
 we will
outline our results and suggest
possible extension of this work.
\section{The effective action for non-BPS
Dp-brane in  Dk-brane background}\label{second}
In this section we will analyse 
the motion of a
non-BPS Dp-brane near  the stack of 
coincident and static Dk-branes 
using the tachyon effective action proposed in
\cite{Sen:1999md,
Bergshoeff:2000dq,Garousi:2000tr,Kluson:2000iy}.
The metric, the dilaton $(\Phi)$, and the R-R
field (C) for  a system of $N$ coincident 
Dk-branes is given by
\footnote{In what follows we consider
the background Dk-branes to be static
and ignore massive closed string modes. 
These modes become relevant at $r\ll l_s$. 
But here we assume $l_s^{7-k}\ll r^{7-k}
\ll Ng_sl_s^{7-k}$. Then the DBI action
is valid in this domain and we can study
the dynamics of branes in terms of
effective action.}
\begin{eqnarray}\label{Dkbac}
g_{\alpha \beta}=H_k^{-\frac{1}{2}}\eta_{\alpha\beta}
\ , g_{mn}=H_k^{\frac{1}{2}}\delta_{mn} \ ,
(\alpha, \beta=0,1,\dots,k \ , 
m, n=k+1,\dots,9) \ , \nonumber \\
e^{2\Phi}=H_k^{\frac{3-k}{2}} \ , 
C_{0\dots k}=H_k^{-1} \ , H_k=1+\frac{\lambda}
{r^{7-k}}  \ , \lambda=Ng_sl_s^{7-k}
\ , \nonumber \\
\end{eqnarray}
where $H_k$ is a harmonic function of $N$ 
Dk-branes satisfying the Green 
function equation in
the transverse space. 

Now let us consider
a non BPS Dp-brane with $p<k$ 
that is inserted
in the background (\ref{Dkbac}) with its
spatial section   stretched
in directions $(x^1,\dots,x^p)$. We will  label
the worldvolume of the non-BPS 
Dp-brane by $\xi^{\mu} ,
\mu=0,\dots,p$ and use reparametrization invariance
of the worldvolume of the Dp-brane 
to set $\xi^{\mu}=x^{\mu}$.
The position of 
the D-brane in the transverse 
directions $(x^{p+1},\dots,x^9)$
gives to rise  to scalar fields on the worldvolume
of D-brane, $(X^{p+1}(\xi^{\mu}),\dots X^9(\xi^{\mu}))$. 
We also restrict ourselves
to the case of homogeneous modes on
the worldvolume of non-BPS Dp-brane.
 Then
the non-BPS Dp-brane effective action
\footnote{This is the form of the
tachyon effective action that was
proposed in \cite{Kutasov:2003er,Niarchos:2004rw}.}
takes the form 
\begin{equation}\label{aTR}
S=-\tau_p \int
dt \sqrt{F}H_k^{\frac{k-p-4}{4}}
\sqrt{1-\dot{Y}^i
\dot{Y}^i-H_k\dot{X}^m\dot{X}^m-
l_s^2H_k^{\frac{1}{2}}
F\dot{T}^2}\equiv -\int dt \mathcal{L} \ ,
\end{equation}
where $\tau_p$ is a tension of the non-BPS Dp-brane
 and where
$F$ is defined as
\begin{equation}
F=\frac{1}{1+\frac{T^2}{2}} \ .
\end{equation}
From here on we reserve 
the coordinates
$X^m$ with ($m=k+1,\dots,9$) 
for dimensions
transverse to Dk-brane and the 
coordinates $Y^i$ with ($i=p+1,\dots,k$) 
for those dimensions 
transverse
to non-BPS Dp-brane, but parallel
to the Dk-brane. 
Finally
in (\ref{aTR}) we have striped off the 
volume factor of the Dp-brane
 spatial section $V_p$.

To simplify the problem further 
we use the  manifest rotation invariance in
transverse $R^{9-k}$ plane. 
Then we can  restrict ourselves
to the motion in the $(x^8,x^9)$ plane where
we introduce the cylindrical
coordinates
\begin{equation}
X^8=R\cos\theta \ ,
X^9=R\sin\theta \ .
\end{equation}

In this paper we will be interested in the region
of the field theory space when non-BPS Dp-brane
is near  the stack of $N$ Dk-branes
($\frac{\lambda}{R^{k-7}}\gg 1$) 
and when  the tachyon is large
($T^2\gg 1$). In this approximation
the  
non-BPS Dp-brane effective action
takes the form
\begin{equation}\label{aTRl}
S=-\tau_p\sqrt{2}
 \lambda^{\frac{k-p-4}{4}}
\int
dt \frac{1}{TR^{\frac{(7-k)(k-p-4)}{4}}}
\sqrt{1-\dot{Y}^i
\dot{Y}^i-\frac{\lambda}{R^{7-k}}(\dot{R}^2
+R^2\dot{\theta}^2)-l_s^2\frac{
2\sqrt{\lambda}}{T^2R^{\frac{7-k}{2}}}\dot{T}^2}
\ .
\end{equation}
Now we come to the main point of
this paper, namely to the definition of
the scaling transformation.   
Let us demand that the action 
(\ref{aTRl}) should be   invariant 
under following 
transformations
\begin{eqnarray}\label{trag}
t'=\Gamma^{\alpha}t \ ,
T'(t')=\Gamma^{\beta}T(t)\ , 
R'(t')=\Gamma^{\gamma}R(t) \ , \nonumber \\
\theta'(t')=\Gamma^{\delta}\theta(t) \ ,
Y'^i(t')=\Gamma^{\epsilon}Y^i(t) \ , 
\lambda=\Gamma^{\omega}\lambda
 \ , \nonumber \\
\end{eqnarray}
where $\Gamma$ is free 
parameter of this scaling transformation
and where  $\alpha,\beta,\gamma,\delta,
\omega,\epsilon$ will be determined
from the requirement that the  
action (\ref{aTRl}) has to be invariant 
with respect to (\ref{trag}). 
For example, if we demand that
\begin{equation}\label{dt}
dt' \frac{\lambda'^{\frac{
k-p-4}{4}}}{T'R'^{\frac{(7-k)(k-p-4)}{4}}}
=dt \frac{\lambda^{\frac{
k-p-4}{4}}}{TR^{\frac{(7-k)(k-p-4)}{4}}}
\end{equation}
we get 
\begin{equation}\label{betaal}
\alpha-\beta-\frac{(7-k)(k-p-4)}{4}\gamma
+\frac{(k-p-4)}{4}\omega
=0  \ .
\end{equation}
If we proceed in the same way
with all other terms in (\ref{aTRl})
we obtain following set of
 equations 
\begin{eqnarray}
\omega-2\alpha=(5-k)\gamma \ , 
\nonumber \\
\omega+2\delta-2\alpha=(7-k)\gamma
  \ ,\nonumber \\
 \omega-4\alpha=(7-k)\gamma
\nonumber \\
\epsilon=\alpha \ . \nonumber \\
\end{eqnarray}
From  the first three equations
we get  
\begin{equation}
\delta=0 \ ,\gamma=-\alpha \ ,
\omega=(k-3)\alpha
\ . 
\end{equation}
Inserting this result into 
(\ref{betaal}) we obtain
\begin{eqnarray}
\beta=\alpha(k-p-3) \ . \nonumber \\
\end{eqnarray}
The result of this
analysis is that
all parameters are expressed
through $\alpha$. Since different
$\alpha$ correspond to different
$\Gamma$ we are free to 
pick one particular value of
$\alpha$.
If we  choose 
 $\alpha=-1$ 
the
scaling transformations take
the form 
\begin{eqnarray}\label{st}
t'=\Gamma^{-1}t \ , 
T'(t')=\Gamma^{-(k-p-3)}T(t) \ ,
R'(t')=\Gamma R(t) \ , \nonumber \\
\theta'(t')=\theta(t) \ , 
Y'^i(t')=\Gamma^{-1}Y^i(t) \ , 
\lambda'=\Gamma^{3-k}\lambda \ .
\nonumber \\
\end{eqnarray}
As we have claimed  in introduction
the action (\ref{aTRl})  is generally 
 invariant under (\ref{st}) on condition
that parameter $\lambda$  
 transforms  
as well
\footnote{One
exception is the case of D3-brane background
that was extensively studied in
 \cite{Kluson:2005qx}.}.
It then follows that
the generator of these transformations
 is not conserved. This can
be seen   as follows.  let us 
 consider following
transformations  
\begin{eqnarray}\label{trang}
t'=t+\omega(t)\epsilon \ , \nonumber \\
\Phi_i'(t')=\Phi_i(t)+\Omega_i(t)\epsilon 
\nonumber \\
m_b'=m_b+\beta_b(t)\epsilon \ , \nonumber \\
\end{eqnarray}
where $\Phi_i$ are dynamical
variables and $m_b$ are
parameters  in the  action 
\begin{equation}\label{actgen}
S=-\int dt \mathcal{L}(\Phi_i,
\dot{\Phi}_i,m_b) \ , 
\end{equation}
 and
where  we have introduced the infinitisemal
parameter $\epsilon \ll 1$.
Now we demand that the action
(\ref{actgen})
is invariant under 
(\ref{trang}) in the following
sense
\begin{equation}
\int_{t_1}^{t_2} dt \mathcal{L}\left(t,\Phi_i(t),
\frac{d\Phi_i}{dt},m_b\right)=
\int_{t_1'}^{t_2'}
 dt'\mathcal{L}\left(t',\Phi'_i(t'),
\frac{d\Phi'_i}{dt'},m'_b\right) \ .
\end{equation}
If we insert (\ref{trang}) into
the right hand side of this
expression and
perform  the Taylor expansion
with respect to  $\epsilon$
we get
\begin{eqnarray}
\int_{t_1}^{t_2} dt\left[
\frac{d}{dt}\left(\omega (\mathcal{L}-
\dot{\Phi}\frac{\delta \mathcal{L}}
{\delta \dot{\Phi}_i})+\Omega_i\frac{\delta
\mathcal{L}}{\delta \dot{\Phi}_i}\right)
+(\Omega_i-\omega\dot{\Phi}_i)
\left(\frac{\delta \mathcal{L}}{\delta
\Phi_i}-\frac{d}{dt}\left(\frac{\delta
\mathcal{L}}{\delta \dot{\Phi}_i}\right)
\right)\right]=-
\int_{t_1}^{t_2}
 dt \frac{\partial \mathcal{L}}{\partial
m_b}\beta_b 
\ .
\nonumber \\
\end{eqnarray}
Now the second term on the left
hand side of
the previous equation is 
 equal to zero thanks to the
equations of motion. 
Then comparing terms on the
left and right side of the
upper expression  we  get
\begin{eqnarray}\label{conJ}
\frac{dJ}{dt}=
-\frac{\partial \mathcal{L}}
{\partial m_b}\beta_b \nonumber \\
J\equiv \omega (\mathcal{L}-
\dot{\Phi}_i\frac{\delta \mathcal{L}}
{\delta \dot{\Phi}_i})+\Omega_i\frac{\delta
\mathcal{L}}{\delta \dot{\Phi}_i} \ . \nonumber \\
\end{eqnarray}
This result explicitly shows that
in case when the parameters
in Lagrangian have to vary ($\beta_b\neq 0)$
in order the action (\ref{actgen}) was invariant
under (\ref{trang}) the generator
of given transformation $J$ is not
conserved.

Using the general discussion
given above we can easily find
the equation that determines
 time evolution of the
generator of the 
transformations (\ref{st}). It is
easy to see that their infinitesimal
form implies following values
of $\omega, \Omega_i$ and $m_b$ 
\begin{eqnarray}
\omega(t)=-t \ , \Omega_T=-(k-p-3)T(t)\ ,
 \Omega_R=R(t) \ , \nonumber \\ 
 \Omega_\theta=0 \ ,
\Omega_{Y^i}=-Y^i(t) \ ,   
\beta_\lambda=(3-k)\lambda \ .  \nonumber \\
\end{eqnarray}
Then from 
(\ref{conJ}) we get that
the generator $J\equiv D$
 is equal to
\begin{equation}
D=tH-(k-p-3)P_TT+P_RR-P_iY^i \ , 
\end{equation}
where $P_R,P_T,P_\theta,P_i$ and
$H$ are 
the canonical momenta 
that using the action 
 (\ref{aTRl}) take the form
\begin{eqnarray}
P_R=\frac{\delta \mathcal{L}}
{\delta \dot{R}}=\frac{
\sqrt{2}\tau_p\lambda^{\frac{k-p-4}{2}}
}{TR^{\frac{(7-k)
(k-p-4)}{4}}}\frac{\lambda\dot{R}}{
R^{7-k}\sqrt{(\dots)}} \ , \nonumber\\
P_{\theta}=\frac{\delta \mathcal{L}}
{\delta \dot{\theta}}=
\frac{\sqrt{2}\tau_p\lambda^{\frac{k-p-4}{2}}}
{TR^{\frac{(7-k)
(k-p-4)}{4}}}\frac{\lambda R^2\dot{\theta}}{
R^{7-k}\sqrt{(\dots)}} \ , \nonumber\\
P_T=\frac{\delta \mathcal{L}}
{\delta \dot{T}}=\frac{
\sqrt{2}\tau_p\lambda^{\frac{k-p-4}{2}}}{TR^{\frac{(7-k)
(k-p-4)}{4}}}\frac{l_s^2\sqrt{\lambda}
\dot{T}}{
T^2R^{\frac{7-k}{2}}\sqrt{(\dots)}} \ , \nonumber\\
P_i=\frac{\delta \mathcal{L}}{\delta
\dot{Y}^i}=\frac{
\sqrt{2}\tau_p\lambda^{\frac{k-p-4}{2}}}
{TR^{\frac{(7-k)
(k-p-4)}{4}}}\frac{\dot{Y}^i}
{\sqrt{(\dots)}} \ . \nonumber \\
\end{eqnarray}
Finally,  the Hamiltonian is
equal to
\begin{equation}\label{Hcan}
H=\sqrt{V^2
+\frac{P_R^2R^{7-k}}{\lambda}
+
\frac{P^2_\theta R^{7-k}}
{\lambda R^2}+P^2_i
+
\frac{P^2_TT^2R^{\frac{7-k}{2}}}
{2l_s^2\sqrt{\lambda}}} \ , 
\end{equation}
where $V^2$ is defined as
\begin{equation}\label{V2}
V^2=\tau_p^2\frac{2}{T^2}\left(
\frac{\lambda}{R^{7-k}}\right)^
{\frac{k-p-4}{2}} \ . 
\end{equation}
Using (\ref{Hcan}) we obtain 
following differential
equations  for $R,\theta, Y^i$
and $T$ 
\begin{eqnarray}\label{dotR}
\dot{R}=\frac{\partial H}{\partial P_R}
=\frac{P_RR^{7-k}}{\lambda E} \ ,
\nonumber \\
\dot{\theta}=
\frac{\partial H}{\partial P_{\theta}}
=\frac{P_\theta R^{7-k}}{\lambda R^2 E} \ ,
\nonumber \\
\dot{T}=\frac{\partial H}
{\partial P_{T}}=
\frac{T^2R^{\frac{7-k}{2}}}
{2l_s^2\sqrt{\lambda}E} \nonumber \\
\dot{Y}^i=\frac{\partial H}
{\partial P_i}=
\frac{P_i}{E} 
 \ , \nonumber \\
\end{eqnarray}
where we have used the fact that Hamiltonian
is conserved and equal to energy $E$. 
It turns out also that it
is useful to  express  the term 
$\frac{\delta \mathcal{L}}{\delta \lambda}$
as a function of canonical
variables 
\begin{eqnarray}
\frac{\delta \mathcal{L}}
{\delta \lambda}=-\frac{E}{4\lambda}
(-2+\frac{(k-p-2)}{E^2}
(V^2+P_i^2)) \ .
\nonumber \\
\end{eqnarray}

Finally, 
in order to 
simplify the 
resulting expressions we 
will  consider from now on 
all momenta $P_i$ to be equal to zero.
 Then
the equation that determines
time evolution  of $D$ takes 
the form
\begin{equation}\label{Dta}
\frac{dD}{dt}=-\lambda(3-k)\frac{\delta
\mathcal{L}}{\delta \lambda}=
\frac{(k-3)E}{2}\left(1-\frac{V^2}{2E^2}
\right) \ .
\end{equation}
It is very difficult to solve
this equation in the full generality.
On the other hand we see  that
the term on the right hand side
will be time independent if
 $V^2=0$ 
and  the integration
of (\ref{Dta}) 
is then trivial. As follows
from (\ref{V2})
this condition can be ensured
 when $T=\pm\infty$. 
In
the next section 
we will study this situation in more detail.
\section{Non-BPS Dp-brane in
its stable vacuum}\label{third}
In this section we will consider
the situation when the tachyon
is static ($P_T=0$) and 
reaches  its global minimum
($T=\pm
\infty$)
\footnote{More general situations
when $P_T\neq 0$ will be considered
in the next publication.}.
 Since now $V^2=0$  
the equation (\ref{Dta})
can be easily integrated 
with the result
\begin{equation}
D=\frac{E(k-3)}{2}t+D_0
 \ ,
\end{equation}
or alternatively
\begin{equation}
D_0=tE\frac{(5-k)}{2}+P_RR \ ,
\end{equation}
where $D_0$ is the value
of the charge $D$ at time $t=0$.
This equation allows us to
express $P_R$ as
a function of $R$ and $t$
so that
\begin{equation}
P_R=\frac{1}{R}\left(
D_0-\frac{(5-k)tE}{2}\right) \ .
\end{equation}
Then  we can easily find
the time dependence of $R$ 
using the first  equation in
(\ref{dotR})
\begin{equation}\label{Rk}
\dot{R}=\frac{P_R R^{7-k}}
{\lambda E}=\frac{1}{\lambda}
\left[\frac{D_0}{E}-\frac{(5-k)}{2}t\right]
R^{6-k}
\end{equation}
that for $k\neq 5$ has the solution 
\begin{equation}\label{Rks}
R^{k-5}=\frac{k-5}{\lambda}
\left[\frac{D_0}{E}t-\frac{(5-k)}{4}
t^2\right]+C \ . 
\end{equation}
 The case $k=5$ will be
analysed separately. Explicitly, 
we obtain:
\begin{itemize}
\item {\bf k=6}

In this case the time
dependence of $R$  given
in (\ref{Rks}) takes the form
\begin{equation} 
R=\frac{1}{\lambda}
\left[\frac{D_0}{E}t+\frac{1}{4}
t^2\right]+R_0 \ , 
\end{equation}
where $R_0$ is the radial distance of
a non-BPS Dp-brane at time $t=0$. 
Note that $R_0$ is related to the
 initial momentum and to the $D_0$ through
the relation $D_0=P_{R}(0)R_0$.
Remarkably, even 
if the tachyon is in its 
stable minimum $T=\pm \infty$
the dynamics of the 
 radial mode is still nontrivial
with the following physical picture:
Non-BPS Dp-brane moves
from the initial position 
$R_i$, that obeys $\frac{\lambda}{R_i}\gg 1$,
 towards to the stack of $N$ D6-branes
until it reaches its turning point
\footnote{As usual  the
turning point is 
the point where the radial velocity of 
Dp-brane is equal to zero: $\dot{R}=0$.}
at $t_T=-\frac{2D_0}{E}$.  Then 
it moves outwards the stack of $N$
D6-branes until it reaches the
region where the approximation
$\frac{\lambda}{R}\gg 1$ ceases
to be valid.  Note that
this result is consistent with
the general discussion performed
in \cite{Kluson:2005qx}.
\item {\bf k=5}

For $k=5$ we obtain  from
(\ref{Rk})
\begin{equation}
\ln R=\frac{1}{\lambda}
\frac{D_0}{E}t+\ln C
\end{equation}
that implies  
\begin{equation}\label{Rk5}
R=R_0 e^{\frac{D_0}{\lambda E}t}\ , 
\end{equation}
where 
$D_0=P_R(0)R_0 \ ,
E=\sqrt{\frac{P^2_R(0)R^2_0}{\lambda}+
\frac{P^2_\theta}{\lambda}}$.
We see that 
for $D_0>0$ the probe 
Dp-brane leaves the worldvolume
of $N$ D5-branes at $t=-\infty$
and moves outwards. After
finite time  it reaches the region where
the approximation of small $R$
ceases to be valid.  
On the other hand for $D_0<0$ we
should consider the positive $t$ only. 
Then Dp-brane starts to move from
its distance $R_0, \frac{\lambda}{R_0}\gg 1$
at time $t=0$ towards to the stack
of $N$ D5-brane that
it reaches at $t=\infty$. 
\item $\mathbf{ k<5}$

In  these cases the solution (\ref{Rks}) can
be written as 
\begin{equation}\label{Rks5}
\frac{\lambda}{R^{5-k}}=
\frac{(5-k)^2}{4}
t^2-(5-k)\frac{D_0}{E}t+
\frac{\lambda}{R^{5-k}_0} \ ,
\end{equation}
where again $R_0$ is the radial
position of a Dp-brane  at $t=0$ and
where $D_0$ and $E$ are equal to
\begin{equation}
D_0=P_R(0)R_0 \ ,E=
\sqrt{\frac{P_R^2(0)R^{7-k}_0}{\lambda}+
\frac{P^2_\theta R^{5-k}_0}{\lambda}} \ . 
\end{equation}
Now the physical picture is different from
the case of D6 and D5-brane background. 
Namely, from (\ref{Rks5}) it is easy
to see that $R$  blows up for finite
$t$ which happens at
\begin{equation}
t_\pm=\frac{2(5-k)}{(5-k)^2}
\left[\frac{D_0}{E}\pm\sqrt{
\frac{D_0^2}{E^2}-\frac{\lambda}{R_0^{5-k}}}
\right] \ .
\end{equation}
Since by presumption we work in
the regime where $\frac{\lambda}{R^{7-k}}\gg 1$
we demand that $R$ is finite for
all $t$.
 This
requirement implies
that
\begin{equation}
\frac{D_0^2}{E^2}-\frac{\lambda}{R_0^{7-k}}<0
\Rightarrow 
\frac{P^2_\theta R_0^{3-k}(0)}
{P^2_R}>0 
\end{equation}
that is always obeyed for $P_\theta\neq 0$.
In this case  we obtain following physical picture.
An unstable
  Dp-brane leaves the stack of
$N$ Dk-branes at $t=-\infty$ and reaches its
turning point at $t_T=\frac{2D_0}
{(5-k)E}$  at the distance
\begin{eqnarray}
R(t_T)=R_0\left[1+\frac{P^2_R(0)}{P^2_\theta}
R_0^2\right]^{\frac{1}{5-k}} \ . 
 \nonumber \\
\end{eqnarray}
After the crossing its
 turning point  
Dp-brane starts  to move
towards to the stack of Dk-branes
which  reaches at the
 $t=\infty$.  
\end{itemize}
In this section we have studied
the dynamics of the non-BPS Dp-brane
in the near horizon region of
 $N$ Dk-branes  when
the tachyon field reaches its
stable minimum. 
We have shown that
in this case there is still 
nontrivial
dynamics present. We mean
that the radial mode now describe
the collective motion of the
gas of highly excited closed strings
that  arises in the process of
 the tachyon condensation.
We will give more comments to this
suggestion in the concussion.
\section{Dynamics of the  BPS Dp-brane
in the near horizon region of $N$
Dk-branes and broken scaling symmetry}
\label{fourth}
This section is devoted to the study
of the dynamics of the BPS Dp-brane
probe near the stack of $N$ Dk-branes
using the scaling symmetry of the
Dp-brane effective action. When we
again restrict ourselves to the
homogeneous fields on the worldvolume
of Dp-brane and to the case of the motion
in the transverse $(x^8,x^9)$ plane
then the action for the BPS Dp-brane
in the near horizon region of
 $N$ Dk-branes
 takes the form
\begin{eqnarray}\label{RA}
S=-\int dt V\sqrt{1-\dot{Y}^i
\dot{Y}^i-\frac{\lambda}{R^{7-k}}(\dot{R}^2
+R^2\dot{\theta}^2)}=
-\int dt \mathcal{L} \ , \nonumber \\
\end{eqnarray}
where we have striped off the 
volume
factor $V_p$ and we have introduced
the potential 
\begin{equation}
V= \frac{T_p \lambda^{\frac{k-p-4}{4}}}
{R^{\frac{(7-k)(k-p-4)}{4}}}
\ . 
\end{equation}
Note also that now $T_p$ is a
 tension of a BPS Dp-brane that is
related to the tension of a 
non-BPS Dp-brane
$\tau_p$ as 
$\tau_p=\sqrt{2}T_p$. 
 Finally, the meaning of
$Y^i$  is the same as in
the case of non-BPS Dp-brane studied
in previous sections. 

 Let us now demand that the action
(\ref{RA}) is invariant under
following transformations:
\begin{equation}\label{scbpsg}
t'=\Gamma^{\alpha}t \ ,
R'(t')=\Gamma^{\gamma}R(t) \ ,
\theta'(t')=\Gamma^{\delta}\theta(t) \ ,
\lambda=\Gamma^{\omega}\lambda \ ,
Y'^i(t')=\Gamma^\beta Y^i(t) \ .
\end{equation}
By comparing the original and
transformed action we obtain 
following set of equations
for parameters $\alpha,\delta,
\omega,\beta$
\begin{eqnarray}\label{rel}
\frac{(k-p-4)}{4}\omega+
\alpha-\frac{(7-k)(k-p-4)}{4}\gamma=0 \ ,
\nonumber \\
\omega+2\gamma-(7-k)\gamma-2\alpha=0 \ ,
\nonumber \\
\omega+2\gamma-(7-k)\gamma
-2\alpha+2\delta=0 \ ,
\nonumber \\
\beta-\alpha=0 \ . \nonumber \\
\end{eqnarray}
From the last three 
 equations we immediately
get that $\delta=0$ and also
\begin{equation}\label{ogal}
\omega+(k-5)\gamma-2\alpha=0
\ . 
\end{equation}
When we insert this relation into 
the first equation in (\ref{rel})
we obtain the equation 
 \begin{equation}\label{rel1}
(k-p-2)\omega+[2(k-5)-(7-k)(k-p-4)]\gamma=0
\end{equation}
that for $(k-p-2)\neq 0$ has the solution
\begin{equation}\label{relsol}
\omega=\frac{1}{(k-p-2)}
[(7-k)(k-p-4)-2(k-5)]\gamma  \ .
\end{equation}
For $(k-p-2)=0$ the equation
 (\ref{rel1})  implies  
$\gamma=0$ and 
(\ref{ogal}) that 
$\omega=2\alpha$.
Then for  $k=p+2$ and for $\alpha=-1$
the scaling transformations
(\ref{scbpsg})
 take the form
\begin{equation}
R'(t')=R(t) \ , 
\lambda'=\Gamma^{-2}\lambda \ ,
t'=\Gamma^{-1}t \ , Y'^i(t')=
\Gamma^{-1}Y^i(t)  \ . 
\end{equation}
One can now determine the 
generator of given transformation 
exactly in the same way as in previous
sections with the result
\begin{equation}\label{Dbps}
D=tH-P_iY^i
\end{equation}
whose time evolution is governed
by equation
\begin{equation}\label{Dt1}
\frac{dD}{dt}=2\lambda
\frac{\delta \mathcal{L}}
{\delta \lambda}=E-\frac{P_i^2}{E} \ ,
\end{equation}
where we have used that
\begin{eqnarray}\label{deltaSl}
\frac{\delta \mathcal{L}}
{\delta \lambda}=-\frac{E}{4\lambda}
(-2+\frac{2P_i^2}{E^2}+\frac{(k-p-2)}{E^2}
V^2) \ .
\nonumber \\
\end{eqnarray}
On the other hand the explicit
time derivative of $D$ given
in (\ref{Dbps}) is equal to
\begin{equation}
\frac{dD}{dt}=
H-P_i\dot{Y}^i=
E-\frac{P_i^2}{E} \ ,
\end{equation}
as a consequence
of the fact that
$H$ and $P_i$ are conserved.
These results then confirm 
the validity of  (\ref{Dt1}).
However we also see that
the integration of (\ref{Dt1})
is trivial and does not simplify
the solution of the equation of
motion for $R$ either. Then in order
to analyse  the dynamics of the radial mode 
we should use the classical
 Hamiltonian treatment that was however
extensively disused   previously in
\cite{Panigrahi:2004qr,Burgess:2003mm}.

Let us now consider another
possibility   when $k-p-4=0$. 
Now looking at the equations
(\ref{rel}) we see that
the first equation here 
implies that $\alpha=0$ and
consequently $\beta=0$. 
Then from the second and third
equations in (\ref{rel})
  we obtain
\begin{equation}
\omega+(k-5)\gamma=0 \ . 
\end{equation}
Since in our convention the
radial mode scales as $R\rightarrow
\Gamma R$ we take  $\gamma=1$.
Then  we obtain
following transformation rules:
\begin{equation}
R'(t')=\Gamma R(t) \ ,
t'=t \ ,
\lambda'=\Gamma^{-(k-5)}\lambda \ . 
\end{equation}
Now it is simple exercise to
determine the corresponding
generator of given transformations
\begin{equation}\label{dprb}
D=
P_RR
\end{equation}
that obeys the equation
\begin{eqnarray}\label{dotDb}
\frac{dD}{dt}
=\frac{(k-5)E}{2}
\left(1-\frac{P_i^2}{E^2}-
\frac{V^2}{E^2}\right) \ .  \nonumber \\
\end{eqnarray}
Note that for $(k-p-4)=0$ the
potential term $V$ equal to $V=T_p$. 

Let us now explicitly solve the
equation of motion for $R$ using
the  equation (\ref{dotDb}).
Since the Hamiltonian
of the probe Dp-brane takes the form
\begin{equation}
H=\sqrt{V^2+\frac{P^2_RR^{7-k}}{\lambda}
+\frac{P^2_\theta R^{5-k}}{\lambda}
+P^2_i}  \ 
\end{equation}
 the equation of motion
for $R$ is equal to 
\begin{equation}\label{dotRbps}
\dot{R}=\frac{\partial H}{
\partial P_R}=\frac{P_R R^{7-k}}
{\lambda E} \ . 
\end{equation}
Since on the right hand side  of the
equation (\ref{dotDb})
we have factors
that do not depend on time 
its integration
is straightforward and 
gives 
\begin{equation}\label{Dtb}
D=P_RR=\frac{(k-5)Et}{2}
\left(1-\frac{T_p^2}{E^2}\right)+D_0 \ , 
\end{equation}
where in order to
simplify the notation we
 have considered 
the situation when
all $P_i$ are equal to zero. 
If we use (\ref{Dtb}) to 
express $P_R$ as function
of $t$ and $R$ we get
\begin{equation}
P_R=\frac{1}{R}
\left(\frac{(k-5)Et}{2}
\left(1-\frac{T_p^2}{E^2}\right) 
+D_0\right) \ . 
\end{equation}
Then inserting the upper expression
into (\ref{dotRbps})
we  obtain 
\begin{equation}
\label{dotRb}
dRR^{k-6}=
\frac{1}{\lambda}
\left(\frac{(k-5)t}{2}
\left(1-\frac{T_p^2}{E^2}\right)
+\frac{D_0}{E}
\right)dt
 \ 
\end{equation}
that for $k\neq 5$ has the 
solution
\begin{equation}\label{rbpsg}
R^{k-5}=\frac{(k-5)}{\lambda}
\left[\frac{(k-5)}{4}t^2
\left(1-\frac{T_p^2}{E^2}\right)
+\frac{D_0}{E}t
\right]+R_0^{k-5} \ .
\end{equation}
Note that for $E=T_p$  that corresponds
to $P_R=P_\theta=0$ we have $R=R_0$.
This fact manifestly demonstrates the
supersymmetric nature of the  configuration
consisting of the Dk-branes
+D(k-4)-branes that do not move.
Now we will discuss the
solution with $T_p\neq E$
in more details. 
\begin{itemize}
\item {\bf k=6}

In this case the solution (\ref{rbpsg})
is equal to 
\begin{equation}\label{k6d}
R=\frac{1}{\lambda}
\left[\frac{1}{4}t^2\left(1-\frac{T_p^2}{E^2}\right)+
\frac{D_0}{E}t
\right]+R_0 \ .
\end{equation}
We  see that the probe 
D2-brane moves initially towards
to the stack of 
$N$ D6-branes reaches its
turning point at $t_T=-
\frac{2D_0}{E}
\left(1-\frac{T_p^2}{E^2}\right)^{-1}$
and then   it moves outwards.
Of course this solution
is only valid for such a time interval
when we have $\frac{\lambda}{R}\gg 1$. 
Note also that this result is
in agreement with the general
analysis performed in 
\cite{Burgess:2003mm}.
\item {\bf k=5}

For $k=5$ the differential
equation (\ref{dotRb}) takes
the form 
\begin{equation}
\dot{R}=\frac{D}{\lambda E}R
\end{equation}
that has the solution
\begin{equation}
R=R_0e^{\frac{D}{\lambda E}t} \ .
\end{equation} 
We see that the character
of the motion of the 
probe D1-brane 
depends on the sign of
$D=P_R(0)R_0$. For
$D<0$  the probe D1-brane
moves towards  the 
 stack of D5-branes
and reaches it at the 
asymptotic future.  On the other
hand the case 
 $D>0$ corresponds to the
emission of D1-brane from
the worldvolume of $N$ D5-branes
at far past ($t=-\infty)$
that then moves outwards and
after some time reaches the
region where the near horizon
approximation breaks down. 
 For $D=0$ the radial
coordinate $R$ is constant.  
\item {\bf k=4}

In this case (\ref{rbpsg})
is equal to
\begin{equation}
\frac{1}{R}=
\frac{1}{\lambda}
\left[\frac{t^2}{4}
\left(1-\frac{T_p^2}{E^2}\right)
-\frac{D_0}{E}t\right]+
\frac{1}{R_0} 
\end{equation}
with the following 
 physical picture.
The probe D0-brane leaves the
stack of $N$ D4-branes at $t=-\infty$ and
moves outwards until it reaches its
turning point at
$t_T=\frac{2D_0}{E}
\left(1-\frac{T_p^2}{E^2}\right)^{-1}$.
Then it moves back and approaches 
the worldvolume of the D4-branes at
$t=\infty$.
\item $\mathbf k<4$
In this case one can easily see
that there is not any dynamical
probe Dp-brane since now 
$p=k-4<0$. 
\end{itemize}
As a final case we will
consider the situation when $k=6$ and
$p=0$.  Then from (\ref{relsol}) we
get that $\omega=0$ and from the
equations in (\ref{rel}) we obtain
that $\gamma=2\alpha$. 
As a result we obtain following
transformation rules for $R,t,Y^i$
and $\lambda$
\begin{equation}
t'=\Gamma^{-1}t \ ,
R'(t')=\Gamma^{-2}R(t) \ ,
Y'^i(t')=\Gamma^{-1}Y^i(t) \ ,
\lambda'=\lambda \ ,
\end{equation}
where we have taken $\alpha=-1$. 
Then the generator of these
transformations is 
\begin{equation}\label{Dk60}
D=tH-P_RR-P_iY^i \ . 
\end{equation}
which is, thanks to the fact
that $\lambda'=\lambda$, conserved. 
Using (\ref{Dk60}) we can again
express  $P_R$ as a  function
of $R,H=E$ and $D$
\begin{equation}\label{Prk60}
P_R=\frac{1}{R}[tE-D] \ ,
\end{equation}
where  we again consider the
case  when $P_i=0$. 
Inserting this relation to the
differential equation for $R$ we
get
\begin{equation}
\dot{R}=\frac{P_R R}{\lambda E}=
\frac{tE-D}{\lambda E}
\end{equation}
that has the solution
\begin{equation}
R=\frac{1}{2\lambda}t^2-\frac{D}{\lambda E}t+
R_0 \ .
\end{equation}
Now we have  the following
picture of the time dependent
dynamics of the probe 
D0-brane in the near
horizon region of $N$ D6-branes
\footnote{In order to avoid the
coupling of the probe D0-brane to
the Ramond-Ramond field we consider
the situation when $P_\theta=0$. 
For discussion of the more
general case when $P_\theta\neq 0$, see 
\cite{Burgess:2003mm}.}. 
The D0-brane starts to move
towards to  the stack 
of $N$ D6-brane at same time
$t_i<0$ that is chosen 
in such a way to ensure
the validity of the near
horizon region approach. Then D0-brane
reaches its 
turning point
at $t_T=\frac{D}{E}$ 
where the direction of its motion changes
and it again moves outwards
until it reaches the region where
the near horizon approximation
breaks down.
\section{Conclusion}\label{fifth}
This paper was devoted to the
study of the dynamics of the
non-BPS and BPS Dp-branes in the
near horizon region of the stack
of $N$ Dk-branes based on an existence
of the broken  scaling symmetry of the probe
effective action in this background. 
We have studied the question
 under which conditions
 Dp-brane effective action is invariant
under this form of the symmetry and we have 
argued that this generally occurs in case when
we allow the scaling parameter $\lambda$ to vary
as well.  Then
we have found
the differential equation that  
determines the time evolution of this
generator and we have argued that 
in the case when the tachyon is sitting in
its stable vacuum this equation can be
easily solved. We mean that it
is very interesting that there
exists nontrivial dynamics of 
an unstable Dp-brane even if the
tachyon reaches its stable minimum.
It is seems to us that this 
 situation is closely related to the
emergence 
of the tachyon matter  \cite{Gibbons:2000hf,Sen:2000kd,
Sen:2003bc,
Kwon:2003qn}. Since it was
argued in these papers that 
 the tachyon 
condensation results into  the emergence of the
gas of very massive 
closed strings that are confined to the
worldvolume of the unstable D-brane
\cite{Sen:2003iv,Sen:2003xs,Lambert:2003zr}
then we can presume  that the
radial mode dynamics captures the collective
motion of the gas of these closed strings. 

We have also performed an analysis of
the  probe BPS Dp-brane in 
the near horizon
region of $N$ Dk-branes that was based
on an emergence of the scaling
symmetry. As in the non-BPS Dp-brane
case  we have found 
the generator of this scaling symmetry 
and the  equation 
that governs its time evolution. 
Then we have solved the equation of 
motion of the probe Dp-brane using
the known time dependence of $D$.

We mean that extension of this work
could be performed in several ways.
Firstly, it would be interesting to study
the scaling symmetry for situations
when the worldvolume fields depend on
spatial coordinates as well. Secondly,
it would be certainly  interesting
to study the properties 
the tachyon matter in general background,
especially it seems very interesting
to study the gas of unstable D-branes
in the context of the 
 brane gas cosmology.
We hope to return to these problems in
future. 
\\
\\
{\bf Acknowledgement}
This work was supported by the
Czech Ministry of Education under Contract No.
MSM 0021622409.



\begin{thebibliography}{20}




\bibitem{Sen:2002qa}
A.~Sen,
\emph{``Time and tachyon,''}
Int.\ J.\ Mod.\ Phys.\ A {\bf 18} (2003) 4869
[arXiv:hep-th/0209122].


\bibitem{Sen:2002in}
A.~Sen,
\emph{``Tachyon matter,''}
JHEP {\bf 0207} (2002) 065
[arXiv:hep-th/0203265].


\bibitem{Sen:2002nu}
A.~Sen,
\emph{``Rolling tachyon,''}
JHEP {\bf 0204} (2002) 048
[arXiv:hep-th/0203211].

\bibitem{Sen:2004nf}
A.~Sen,
\emph{``Tachyon dynamics 
in open string theory,''}
arXiv:hep-th/0410103.






\bibitem{Kutasov:2004dj}
D.~Kutasov,
\emph{``D-brane dynamics 
near NS5-branes,''}
arXiv:hep-th/0405058.








\bibitem{Nakayama:2004ge}
Y.~Nakayama, K.~L.~Panigrahi, 
S.~J.~Rey and H.~Takayanagi,
\emph{``Rolling down the throat 
in NS5-brane background: 
The case of electrified
D-brane,''}
arXiv:hep-th/0412038.

\bibitem{Chen:2004vw}
B.~Chen, M.~Li and B.~Sun,
\emph{``Dbrane near NS5-branes: 
With electromagnetic field,''}
arXiv:hep-th/0412022.

\bibitem{Thomas:2004cd}
S.~Thomas and J.~Ward,
\emph{``D-brane dynamics 
and NS5 rings,''}
arXiv:hep-th/0411130.

\bibitem{Bak:2004tp}
D.~Bak, S.~J.~Rey and H.~U.~Yee,
\emph{``Exactly soluble dynamics 
of (p,q) string near 
macroscopic fundamental
strings,''}
arXiv:hep-th/0411099.

\bibitem{Kluson:2004yk}
J.~Kluson,
\emph{``Non-BPS Dp-brane 
in the background of 
NS5-branes on transverse R**3 x
S**1,''}
arXiv:hep-th/0411014.


\bibitem{Kluson:2004xc}
J.~Kluson,
\emph{``Non-BPS D-brane near NS5-branes,''}
JHEP {\bf 0411}, 013 (2004)
[arXiv:hep-th/0409298].

\bibitem{Saremi:2004yd}
O.~Saremi, L.~Kofman and A.~W.~Peet,
\emph{``Folding branes,''}
arXiv:hep-th/0409092.


\bibitem{Kutasov:2004ct}
D.~Kutasov,
\emph{``A geometric interpretation 
of the open string tachyon,''}
arXiv:hep-th/0408073.



\bibitem{Sahakyan:2004cq}
D.~A.~Sahakyan,
\emph{``Comments on D-brane 
dynamics near NS5-branes,''}
JHEP {\bf 0410}, 008 (2004)
[arXiv:hep-th/0408070].


\bibitem{Ghodsi:2004wn}
A.~Ghodsi and A.~E.~Mosaffa,
\emph{``D-brane dynamics in 
RR deformation of NS5-branes 
background and tachyon
cosmology,''}
arXiv:hep-th/0408015.

\bibitem{Panigrahi:2004qr}
K.~L.~Panigrahi,
\emph{``D-brane dynamics in Dp-brane background,''}
Phys.\ Lett.\ B {\bf 601}, 64 (2004)
[arXiv:hep-th/0407134].

\bibitem{Yavartanoo:2004wb}
H.~Yavartanoo,
\emph{``Cosmological solution 
from D-brane motion in NS5-branes
 background,''}
arXiv:hep-th/0407079.

\bibitem{Nakayama:2004yx}
Y.~Nakayama, Y.~Sugawara and H.~Takayanagi,
\emph{``Boundary states 
for the rolling D-branes 
in NS5 background,''}
JHEP {\bf 0407}, 020 (2004)
[arXiv:hep-th/0406173].

\bibitem{Kluson:2005qx}
J.~Kluson,
\emph{``Non-BPS Dp-brane in 
Dk-brane background,''}
arXiv:hep-th/0501010.




\bibitem{Sen:1999md}
A.~Sen,
\emph{``Supersymmetric 
world-volume action for non-BPS D-branes,''}
JHEP {\bf 9910} (1999) 008
[arXiv:hep-th/9909062].

\bibitem{Kluson:2000iy}
J.~Kluson,
\emph{``Proposal for non-BPS D-brane action,''}
Phys.\ Rev.\ D {\bf 62} (2000) 126003
[arXiv:hep-th/0004106].

\bibitem{Bergshoeff:2000dq}
E.~A.~Bergshoeff, M.~de Roo, 
T.~C.~de Wit, E.~Eyras and S.~Panda,
\emph{``T-duality and actions for non-BPS D-branes,''}
JHEP {\bf 0005} (2000) 009
[arXiv:hep-th/0003221].

\bibitem{Garousi:2000tr}
M.~R.~Garousi,
\emph{``Tachyon couplings 
on non-BPS D-branes and 
Dirac-Born-Infeld action,''}
Nucl.\ Phys.\ B {\bf 584} (2000) 284
[arXiv:hep-th/0003122].


\bibitem{Kutasov:2003er}
D.~Kutasov and V.~Niarchos,
\emph{``Tachyon effective 
actions in open string theory,''}
Nucl.\ Phys.\ B {\bf 666} (2003) 56
[arXiv:hep-th/0304045].

\bibitem{Niarchos:2004rw}
V.~Niarchos,
\emph{``Notes on tachyon 
effective actions and Veneziano amplitudes,''}
Phys.\ Rev.\ D {\bf 69} (2004) 106009
[arXiv:hep-th/0401066].

\bibitem{Fotopoulos:2003yt}
A.~Fotopoulos and A.~A.~Tseytlin,
\emph{``On open superstring partition 
function in inhomogeneous rolling tachyon
background,''}
JHEP {\bf 0312}, 025 (2003)
[arXiv:hep-th/0310253].

\bibitem{Sen:2003zf}
A.~Sen,
\emph{``Moduli space of unstable 
D-branes on a circle of critical radius,''}
JHEP {\bf 0403} (2004) 070
[arXiv:hep-th/0312003].



\bibitem{Burgess:2003qv}
C.~P.~Burgess, P.~Martineau, 
F.~Quevedo and R.~Rabadan,
\emph{``Branonium,''}
JHEP {\bf 0306} (2003) 037
[arXiv:hep-th/0303170].

\bibitem{Burgess:2003mm}
C.~P.~Burgess, N.~E.~Grandi, 
F.~Quevedo and R.~Rabadan,
\emph{``D-brane chemistry,''}
JHEP {\bf 0401} (2004) 067
[arXiv:hep-th/0310010].


\bibitem{Sen:2003tm}
A.~Sen,
\emph{``Dirac-Born-Infeld 
action on the tachyon 
kink and vortex,''}
Phys.\ Rev.\ D {\bf 68} (2003) 066008
[arXiv:hep-th/0303057].





\bibitem{Jevicki:1998ub}
A.~Jevicki, Y.~Kazama and T.~Yoneya,
\emph{``Generalized conformal 
symmetry in D-brane matrix models,''}
Phys.\ Rev.\ D {\bf 59} (1999) 066001
[arXiv:hep-th/9810146].

\bibitem{Jevicki:1998yr}
A.~Jevicki and T.~Yoneya,
\emph{``Space-time uncertainty 
principle and conformal symmetry in D-particle
dynamics,''}
Nucl.\ Phys.\ B {\bf 535} (1998) 335
[arXiv:hep-th/9805069].




\bibitem{Gibbons:2000hf}
G.~W.~Gibbons, K.~Hori and P.~Yi,
\emph{``String fluid from unstable D-branes,''}
Nucl.\ Phys.\ B {\bf 596}, 136 (2001)
[arXiv:hep-th/0009061].

\bibitem{Sen:2000kd}
A.~Sen,
\emph{``Fundamental strings 
in open string theory at the tachyonic vacuum,''}
J.\ Math.\ Phys.\  {\bf 42}, 2844 (2001)
[arXiv:hep-th/0010240].

\bibitem{Sen:2003bc}
A.~Sen,
\emph{``Open and closed strings 
from unstable D-branes,''}
Phys.\ Rev.\ D {\bf 68}, 106003 (2003)
[arXiv:hep-th/0305011].

\bibitem{Kwon:2003qn}
O.~K.~Kwon and P.~Yi,
\emph{``String fluid, tachyon matter, 
and domain walls,''}
JHEP {\bf 0309} (2003) 003
[arXiv:hep-th/0305229].




\bibitem{Sen:2003iv}
A.~Sen,
\emph{``Open-closed duality: 
Lessons from matrix model,''}
Mod.\ Phys.\ Lett.\ A {\bf 19} (2004) 841
[arXiv:hep-th/0308068].

\bibitem{Sen:2003xs}
A.~Sen,
\emph{``Open-closed duality at tree level,''}
Phys.\ Rev.\ Lett.\  {\bf 91} (2003) 181601
[arXiv:hep-th/0306137].

\bibitem{Lambert:2003zr}
N.~Lambert, H.~Liu and J.~Maldacena,
\emph{``Closed strings from decaying D-branes,''}
arXiv:hep-th/0303139.











\end{thebibliography}
\end{document}